\documentclass{ws-ijgmmp}

\begin{document}

\markboth{D. Momeni,  R. Myrzakulov}
{   Tolman-Oppenheimer-Volkoff  Equations in  Modified Gauss-Bonnet Gravity}

%
\catchline{}{}{}{}{}
%

\title{Tolman-Oppenheimer-Volkoff  Equations in  Modified Gauss-Bonnet Gravity
}

\author{D. Momeni}

\address{Eurasian International Center for Theoretical Physics and Department of General
\& Theoretical Physics, Eurasian National University, \\ Astana 010008, Kazakhstan\\
d.momeni@yahoo.com }

\author{R. Myrzakulov}

\address{Eurasian International Center for Theoretical Physics and Department of General
\& Theoretical Physics, Eurasian National University, \\ Astana 010008, Kazakhstan\\
rmyrzakulov@gmail.com }

\maketitle

\begin{history}
\received{(Day Month Year)}
\revised{(Day Month Year)}
\end{history}

\begin{abstract}
Based on a stringy inspired Gauss-Bonnet (GB) modification of classical gravity, we constructed a model for neutron stars. We derived the modified forms of Tolman-Oppenheimer-Volkoff (TOV) equations for a generic function of $f(G)$ gravity. The hydrostatic equations remained unchanged but the dynamical equations for metric functions are modified due to the effects of GB term. 
\end{abstract}


\section{Introduction}
Observational data show that we live in an accelerating Universe at large scales \cite{obs1}-\cite{obs3}. This behavior is governed beyound the classical dynamics of solar objects and even beyound the Einstein theory for gravity as a gauge theory,so called as general relativity (GR). So to explain this acceleration expansion at large cosmological scales, we need some kinds of modifications to the classical GR. This approach is called as modified gravity and it is believed that it can solve the problem of acceleration without any need to exotic fluids (for reviews see \cite{Nojiri:2010wj}-\cite{ijgmmp}). One of the simplest modifications of GR is to  replace the Ricci scalar $R$ by an arbitrary function $f(R)$. It is called as $f(R)$ gravity and originally proposed by Buchdahl \cite{Buchdahl} and recently motivated by several authors. In the context of Riemmanian geometry the next higher order correction to $R$ is called as GB term is defined by  $G=R^2-4R_{\mu\nu}R^{\mu\nu}+R_{\mu\nu\lambda\sigma}R^{\mu\nu\lambda\sigma}$. Indeed it is a topological invariant term in four dimensions and if we add it to the Ricci scalar term in Einstein gravitional action, it has no contribution to the equations of motion. But a non minimally coupled form of this topological term induces new features as an alternative for dark energy.  This modified $f(G)$ gravity has been introduced as model for dark energy in \cite{sergei3}.  The GB term plays a very important role in cosmological models as several works have been devoted to the interesting aspects of this term \cite{Capozziello:2014bqa}-\cite{Nojiri:2006je}. In fact this term is inspired from string theory. In low energy limit of the stringy action the first second order term is GB. Also if we study the most general scalar-tensor theory with second order field equation ,we observe this GB term \cite{Horndeski}. \par
From other point of view, dynamics of compact objects is very important problem in astrophysics. One of the most important objects in neutron star. Neutron stars compact objects with radius of order $10   (Km)$\cite{book}. But they are so massive. For a typical neutron star, the mass is about $1.4 M_{*}$ where $M_{*}$ denotes mass of our Sun. Since it is extremely massive and tiny in comparison to our planet,a surface gravity on this star is much higher than Earth. The relative  order of surface gravity of neutron star in comparison  to Earth is about $2\times 10^{11}$. One important property of neutron stars is they can be charged massive objects. But the strength of electromagnetic fields of such stars are about some million times stronger than the one which we have on Earth. In a very close competition with neutron stars,we can have White Dwarfs or black holes. The main difference is in the order of mass of these massive objects. Gravitational collapse is the dominant mechanism in formaion of all these three types of massive objects. \par
In GR several models have been studied for neutron stars. In modified gravity $f(R)$ there are few refrences about neutron stars namely \cite{Astashenok:2014gda}-\cite{Astashenok:2013vza}. But in $f(G)$ gravity there is no work about neutron stars. In this letter we explore TOV equations for a gravitational theory with GB term. We derived the full system of equations of motion for a spherically symmetric static configuration with inhomogenous perfect fluid. We showed that how the equations of motion will be changed due to GB corrections. 

\section{ TOV equations in Modified Gauss-Bonnet gravity}
Let us to start by the gravitational action of a generic GB model as the following  \cite{sergei3}:
\begin{eqnarray}
S=\int d^4x\sqrt{-g}\left[\frac{R}{2\kappa^2}+f(G)\right]+S_m\,\,,\label{S}
\end{eqnarray}
Since this action in defined by commutative connections and in the framework of 
 a Riemannian spacetime, as usual the second order scalar curvature  $R$ is defined by the Ricci scalar. The next higher order correction due to the GB term is denoted by  the algebraic function $f(G)$. We assume that this function is smooth enough to have higher order derivatives $f^{n\geq2}(G)$. It is a necesary condition for linear stability. To include the matter fields we add a matter action 
   $S_m$. The conntribution of this term to the field equations  induces  the energy momentum tensor $T_{\mu\nu}=-\frac{2}{\sqrt{-g}}\frac{\delta S_{m}}{\delta g^{\mu\nu}}$. Furthermore,the gravitational coupling  is defined by   $\kappa^2=8\pi G_N$. Here $G_N$ is the tiny usual classical Newtonian gravitational coupling. In metric formalism of modified gravity when we take metric as dynamical variable , the equations of motion  from (\ref{S}) are written as the following form:
\begin{eqnarray}
R_{\mu\nu}-\frac{1}{2}Rg_{\mu\nu}+8\Big[R_{\mu\rho\nu\sigma}+R_{\rho\nu}g_{\sigma\mu}
-R_{\rho\sigma}g_{\nu\mu}-R_{\mu\nu}g_{\sigma\rho}+R_{\mu\sigma}g_{\nu\rho}\nonumber\\
+\frac{R}{2}\left(g_{\mu\nu}g_{\sigma\rho}-g_{\mu\sigma}g_{\nu\rho}\right)\Big]\nabla^{\rho}\nabla^{\sigma}f_{G}+\left(Gf_{G}-f\right)g_{\mu\nu}=\kappa^2 T_{\mu\nu}\,\,,\label{eom}
\end{eqnarray}
where $f_{GG...}=\frac{d^nf(G)}{dG^n}$ . Further more we adopted the signature of the metric $g_{\mu\nu}$ as $(+---)$ , and consequently  $\nabla_{\mu}V_{\nu}=\partial_{\mu}V_{\nu}-\Gamma_{\mu\nu}^{\lambda}V_{\lambda}$ and $R^{\sigma}_{\;\mu\nu\rho}=\partial_{\nu}\Gamma^{\sigma}_{\mu\rho}-\partial_{\rho}
\Gamma^{\sigma}_{\mu\nu}+\Gamma^{\omega}_{\mu\rho}\Gamma^{\sigma}_{\omega\nu}
-\Gamma^{\omega}_{\mu\nu}\Gamma^{\sigma}_{\omega\rho}$ for the covariant derivative and the Riemann tensor, respectively.  There is an additional conservation law for the matter sector as $\nabla^{\mu}T_{\mu\nu}=0$. 
\par
We suppose that the metric of the neutron star is static-spherically symmetric with coordinates $x^{\mu}=(ct,r,\theta,\varphi)$ in the following form:
\begin{eqnarray}
ds^2=c^2 e^{2\phi}dt^2-e^{2\lambda}dr^2-r^2(d\theta^2+\sin^2\theta d\varphi^2)\label{g}.
\end{eqnarray}
For the matter field, we suppose that the non-zero components of energy momentum tensor are $T_{\mu}^{\nu}=diag(\rho c^2,-p,-p,-p)$. 
We insert (\ref{g}) in (\ref{eom}) and by using the formula given in appendix, the diagonal components of (\ref{eom}) for  $(\mu,\nu)=(ct,ct)$ and $(\mu,\nu)=(r,r) $ are given by the following equations:
\begin{eqnarray}
&&-\frac{1}{r^2}(2r\lambda'+e^{2\lambda}-1)+8e^{-2\lambda}
\Big(f_{GG}(G''-2\lambda'G')+f_{GGG}(G')^2\Big)
\Big(\frac{1-e^{2\lambda}}{r^2}-2(\phi''+\phi'^2)\Big)\\&&\nonumber+(Gf_{G}-f)e^{2\lambda}=\kappa^2\rho c^2 e^{2\lambda}\label{tt}.
\end{eqnarray}
\begin{eqnarray}
&&-\frac{1}{r^2}(2r\phi'-e^{2\lambda}+1)-(Gf_{G}-f)e^{2\lambda}=\kappa^2 p e^{2\lambda}\label{rr}.
\end{eqnarray}
We need an additional equation for metric functions and f(G). We use the 
trace of (\ref{eom}). We obtain:
\begin{eqnarray}
&&R+8G_{\rho\sigma}\nabla^{\rho}\nabla^{\sigma}f_{G}-4(Gf_{G}-f)=-\kappa^2(\rho c^2-p).
\end{eqnarray}
Using metric (\ref{g}) it reads as the follows:
\begin{eqnarray}
&&2\Big(\phi''+\phi'^2-\phi'\lambda'+\frac{2}{r}(\phi'-\lambda')+\frac{1-e^{2\lambda}}{r^2}\Big)\\&&\nonumber
+8 e^{-2\lambda}\Big(\frac{2\phi'}{r}+\frac{1-e^{2\lambda}}{r^2}\Big)\Big(f_{GG}(G''-2\lambda'G')+f_{GGG}(G')^2\Big)+4(Gf_{G}-f)e^{2\lambda}=\kappa^2e^{2\lambda}(\rho c^2-3p)\label{T}.
\end{eqnarray}
The hydrostatic (continuty equation) follows from the equation , $\nabla_{\mu}T_{\nu}^{\mu}$ for $\nu=r$. We obtain:
\begin{eqnarray}
\frac{dp}{dr}=-(p+\rho c^2)\phi'\label{p}.
\end{eqnarray}
The continuty equation is  satisfied identically for $\nu=t$.
Note that when $f(G)=G$ ,then (\ref{tt},\ref{rr}) reduce to the GR equations trivially.\par
To have proceed in TOV equations,
we replace the metric function with the following expression in terms of the gravitational mass function $M=M(r)$:
\begin{eqnarray}
e^{-2\lambda}=1-\frac{2GM}{c^2 r}\Longrightarrow \frac{G dM}{c^2 dr}=\frac{1}{2}\Big[1-e^{-2\lambda}(1-2r\lambda')\Big]\label{dM}.
\end{eqnarray}
The plan is to rewrite (\ref{tt},\ref{rr},\ref{T}) in terms of $\{\frac{dp}{dr},\frac{dM}{dr},\rho\}$ in a dimensionless form. For this purpose,we introduce the following set of the dimensionless parameters:
\begin{eqnarray}
M\to m M_{\star},\ \ r\to r_{g} r,\ \ \rho\to \frac{\rho M_{\star}}{r_{g}^3},\ \ p\to \frac{ p M_{\star} c^2}{r_{g}^3},\ \ G\to\frac{G}{r_g^4}.
\end{eqnarray}
Here $r_{g}=\frac{GM_{\star}}{c^2}=1.47473 (Km)$. From now we use the dimensionless parameters. Continuty equation converts to the following :
\begin{eqnarray}
\frac{dp}{dr}=-(p+\rho )\phi'\label{p2}.
\end{eqnarray}
  But (\ref{dM}) converts to the following:
\begin{eqnarray}
\frac{d\lambda}{dr}=\frac{m}{r^3}\frac{1-\frac{r}{m}\frac{dm}{dr}}{\frac{2m}{r}-1}\label{dm}.
\end{eqnarray}
Using the (\ref{p2},\ref{dm}) in (\ref{rr}) we obtain:
\begin{eqnarray}
\frac{2}{r}\frac{1-\frac{2m}{r}}{p+\rho c^2}+\frac{2m}{r^3}-r_g^2(Gf_{G}-f)=8\pi p\label{eq1}.
\end{eqnarray}
For ($rr$) equation we obtain:
\begin{eqnarray}
&&-\frac{2}{r^2}\frac{dm}{dr}+8r_g^2(1-\frac{2m}{r})^2\Big[f_{GG}\Big(r_g^2G''-r_g^3\frac{2m}{r^2}\frac{1-r\frac{dm}{dr}}{\frac{2m}{r}-1}G'\Big)+r_g^2f_{GGG}G'^2\Big]\\&&\times
\Big[-\frac{2m/r^3}{1-\frac{2m}{r}}+2\frac{d}{dr}\Big(\frac{\frac{dp}{dr}}{p+\rho }\Big)-2\Big(\frac{\frac{dp}{dr}}{p+\rho }\Big)^2\Big]\nonumber+(Gf_{G}-f)=8\pi \rho\label{eq22}.
\end{eqnarray}
And the last equation for trace, in the dimensionless parameters reads:
\begin{eqnarray}
&&2(1-\frac{2m}{r})\Big(-\frac{d}{dr}(\frac{\frac{dp}{dr}}{p+\rho})+(\frac{\frac{dp}{dr}}{p+\rho})^2+\frac{m}{r^3}\frac{1-\frac{r}{m}\frac{dm}{dr}}{\frac{2m}{r}-1}(\frac{\frac{dp}{dr}}{p+\rho})-\frac{2m/r^3}{1-\frac{2m}{r}}\Big)\\&&\nonumber
+8\Big[-\frac{2\frac{dp}{dr}}{r(p+\rho)}-\frac{2m/r^3}{1-\frac{2m}{r}}\Big]\Big[f_{GG}\Big(r_g^2G''-r_g^3\frac{2m}{r^2}\frac{1-r\frac{dm}{dr}}{\frac{2m}{r}-1}G'\Big)+r_g^2f_{GGG}G'^2\Big]\\&&\nonumber
4(Gf_{G}-f)=8\pi (\rho-3p)\label{eq3}.
\end{eqnarray}
The full set of these equations can be used to construct a model for neutron star\cite{Astashenok:2014nua}
.


\section{Conclusion}
Neutron stars are massive,compact astrophysical objects. They can be modelled as spherically symmetric stars with a specific equation of state. In general relativity several models of neutron stars have been investigated. Specially when we adopted a linear equation of state,the model has been studied as quark matter star. In modified gravity of $f(R)$ gravity,this problem has been solved perfectly using an empirical equation of state. In this work we formulated neutron stars in a Gauss-Bonnet modified gravity,so called as $f(G)$ gravity in which G denotes the topological invariant term of four dimensional spacetime. We derived the full set of equations of motion for spherically symmetric metric filled by cosmic fluid. We also derived the modified  Tolman-Oppenheimer-Volkoff  equation. This equation shows the dynamical behavior of the model in terms of the mass,pressure and energy density. This set of  equations is dimensionless. We should specify the model for $f(G)$. A way to find the solutions is to think about  the modified part as a perturbation. So, we are able to perform perturbations for all variables. But a more realistic approach is to solve equations numerically which is beyound our current letter.

\section*{Acknowledgments}
We would like to thank  the anonymous reviewer for enlightening comments related to this work.

\section{Appendices}
In this appendix we present different geometrical quantities as we used in this letter. For metric (\ref{g}) the following non zero components of the symmetric connection are obtained:
\begin{eqnarray}
\Gamma_{12}^1=\phi',\ \ \Gamma_{11}^2=\phi' e^{2\phi-2\lambda},\ \ \Gamma_{22}^{2}=\lambda',\ \ \Gamma_{23}^2=-re^{-2\lambda},\\
\Gamma_{44}^2=-r\sin^2\theta e^{-2\lambda},\ \ \Gamma_{23}^3=\frac{1}{r},\ \ \Gamma_{44}^3=-\sin\theta\cos\theta, \ \ \Gamma_{24}^4=\frac{1}{r},\ \ \Gamma_{34}^4=\cot\theta.
\end{eqnarray}
So, the non vanishing components of Einstein tensor read:
\begin{eqnarray}
G_{11}&=&-e^{2\phi-2\lambda}(\frac{2\lambda'}{r}+\frac{e^{2\lambda}-1}{r^2}),\\
G_{22}&=&-\frac{2\phi'}{r}+\frac{e^{2\lambda}-1}{r^2}\\
G_{33}&=&-re^{-2\lambda}\Big(\phi'-\lambda'+r(\phi''+\phi'^2)-r\phi'\lambda'\Big)\\
G_{44}&=&-\sin^2\theta G_{33}.
\end{eqnarray}
For Riemann tensor we obtain:
\begin{eqnarray}
R_{1212}=e^{2\phi}(\phi'\lambda'-\phi''-\phi'^2),\ \ R_{1313}=-re^{2\phi-2\lambda}\phi',\ \ R_{1414}=sin^2\theta R_{1313},\\
R_{2323}=-r\lambda',\ \ R_{2424}=\sin^2\theta R_{2323},\ \ R_{3434}=\sin^2\theta r^2 e^{-2\lambda}(1-e^{2\lambda}).
\end{eqnarray}
Thus the Ricci scalar is as the following:
\begin{eqnarray}
R=-2e^{-2\lambda}\Big(\phi''+\phi'^2-\phi'\lambda'+2\frac{\phi'-\lambda'}{r}+\frac{1-e^{2\lambda}}{r^2}\Big).
\end{eqnarray}
The GB term reads:
\begin{eqnarray}
&&\frac{-e^{4\lambda}G}{8}=\frac{1}{r^2}\Big[(\phi''+\phi'^2-\lambda'\phi')(e^{2\lambda}-1)+2\phi'\lambda'\Big]
\end{eqnarray}
To derive the hydrostatic equation we use the following equation:
\begin{eqnarray}\nonumber
&&\nabla_{\mu}T^{\mu}_{2}=\partial_{\mu}T^{\mu}_{2}
+\Gamma^{\mu}_{\mu\sigma}T^{\sigma}_{2}-\Gamma^{\sigma}_{\mu 2}T^{\mu}_{\sigma}=0,\   \ \partial_{\mu}(-p\delta^{\mu,2})-p\Gamma_{\mu 2}^{\mu}+p\Gamma^{a}_{a 2}-\rho c^2 \Gamma^{1}_{12}=0,\\&&\nonumber
-\frac{dp}{dr}-(p+\rho c^2)\Gamma^{1}_{12}=0\Longrightarrow \frac{dp}{dr}=-(p+\rho c^2)\phi'.
\end{eqnarray}

\end{document}